\documentstyle[12pt]{article}

\newcommand{\be}{\begin{equation}}
\newcommand{\ee}{\end{equation}}
\newcommand{\bea}{\begin{eqnarray}}
\newcommand{\eea}{\end{eqnarray}}

\newcommand{\un}{\underline}
\newcommand{\ve}{\varepsilon}
\newcommand{\ba}{\begin{array}}
\newcommand{\ea}{\end{array}}
\newcommand{\cchi}{\raisebox{2pt}{$\chi$}}

\begin{document}

\begin{flushright}
UM-TH-97-16, hep-th/9707193
\end{flushright}

\vspace{0.5cm}

\centerline{\Large \bf N=2, D=6 supergravity
with $E_7$ gauge matter}

\vspace{0.5cm} 

\centerline{{\large K.Zyablyuk}\footnote{On leave 
from the Institute of Theoretical and
Experimental Physics, Moscow;\\ e-mail:   
{\tt zyablyuk@vxitep.itep.ru}}}
\centerline{\it Randall Laboratory of Physics, 
University of Michigan, Ann Arbor, MI 48109-1120}

\begin{abstract}
The lagrangian of N=2, D=6 supergravity coupled to 
$E_7\times SU(2)$ vector- and hyper-multiplets is derived. 
For this purpose the coset manifold $E_8/E_7\times SU(2)$,
parametrized by the scalars of the hypermultiplet, is 
constructed. A difference from the case of $Sp(n)$-matter 
is pointed out. This model can be considered as
an intermediate step in the compactification of
D=10 supergravity coupled to $E_8\times E_8$ matter
to four-dimensional model of $E_6$ unification.
\end{abstract}

\section*{Introduction}

Minimal six-dimensional supergravity has N=2
supersymmetries and can be coupled to the vector multiplet
in adjoint representation of arbitrary gauge group.
On the other hand, the hypermultiplet must belong
to certain group to be coupled to the supergravity. 
In particular, the complete lagrangian with all
couplings has been constructed in \cite{NS} for the
case of $Sp(n)\times SU(2)$ gauge group. The group $SU(2)$
acts on two supersymmetry generators while 
$Sp(n)$ transforms only matter fields. The scalars of the 
hypermultiplet parametrize the coset manifold
$Sp(n,1)/Sp(n)\times SU(2)$, which possesses special 
properties, allowing one to construct the supersymmetric 
action.

Other coset manifolds were suggested in \cite{NS,BW}
as candidates for the spaces the hypermatter could 
form. Of particular interest is the largest exceptional
one, namely $E_8/E_7\times SU(2)$. In this paper
we explicitly construct this manifold and present the
lagrangian of the N=2, D=6 supergravity coupled to
the $E_7\times SU(2)$ vector- and hyper-multiplets.
In the rest of the Introduction we argue, that this
model may play important role in the compactification
of the $E_8\times E_8$ heterotic string. 

The compactification of D=10 supergravity, which is
the low-energy limit of superstrings, to 
four-dimensional space-time is rather ambiguous
\cite{CHSW}. The topological structure of the
internal Calabi-Yau manifold is not determined.
Moreover, complexity of six-dimensional spaces
requires many unknown parameters to be introduced
in order to obtain a predictable multigeneration
model in four dimensions.

On the other hand, if one suggests, that at some
intermediate energy-scale between the Plank mass
and the Grand Unification scale the space-time is
effectively six-dimensional, many problems get fixed.
In this case internal four-dimensional manifold
has $SU(2)$ holonomy group and, consequently,
selfdual Ricci tensor; this is $K_3$ with necessity
\cite{GSW}.
Moreover, the vacuum configuration guarantees the
vanishing cosmological constant in six dimensions
\cite{TZ}, even if higher-derivative corrections are 
included. 

After the compactification to six dimensions 
gauge vectors with
four-dimensional index become scalars. They belong
to (248,1)+(1,248) representation of  
$E_8\times {E_8}'$ gauge group and 4=(2,1)+(1,2)
representation of $O(4)\sim SU(2)\times SU'(2)$
Lorentz group of internal manifold. Since
$Tr_{E_8\times {E_8}'}=Tr_{E_8}+Tr_{{E_8}'}$ one can expect
that ${E_8}'$ decouples. It is then natural
\cite{FKP} to pick up the singlet of 
$SU'(2)\times SU''(2)$, where $SU''(2)$ is a subgroup of
$E_8$. The residual group is $E_7\times SU(2)$ and
the scalars belong to its $(56,2)$ representation. They
can form the coset manifold mentioned above.

Nevertheless the compactification
N=1, D=10 $\to$ N=2, D=6 may turn out to be difficult
to perform explicitly. In particular, the vector 
responsible for gauging of $SU(2)$ group is the
ten-dimensional spin-connection $\hat{\omega}_m$,
which is composite field, rather than independent
degree of freedom. But since all couplings of
six-dimensional supergravity are fixed uniquely up 
to two gauge coupling constants $g$ and $g'$, it is
interesting to construct the lagrangian independently. 
Possibly, the consistent consideration of the
compactification may determine these constants too.

The compactification N=2, D=6 $\to$ N=1, D=4 shouldn't
be a problem since all two-dimensional equations of
motion are explicitly solvable. The $E_7$ group is
expected to be broken down to anomaly free 
$E_6$, or further to $SO(10)$ group. One may hope
that chiral compactification \cite{SS} on the 
sphere $S^2$ can solve the problem of mirror generations,
which inevitably present in any real group such
as $E_7$ or $E_8$.

\section*{$E_8/E_7\times SU(2)$ coset manifold}

In order to construct the coset manifold 
$E_8/E_7\times SU(2)$, we need $E_8$ algebra. For this
purpose we use the $E_8\supset E_7\times SU(2)$
decomposition:
$$
248\,=\,(133,1)\,+\,(56,2)\,+\,(1,3)
$$
So the generators of $E_8$ are
$(Y_\Sigma,\,I_s,\,Q_{i\alpha})$,
where $Y_\Sigma$ are $E_7$ generators,
$I_s$ are $SU(2)$ ones and
$Q_{i\alpha}$ are off-diagonal
coset generators. All index notations are given in Table 1.
\begin{table}[b]
\begin{center}
\begin{tabular}{lll}\hline
indices: & values: & representation: \\ \hline\hline
$\Sigma,\,\Lambda,\,\Pi$ & $1,\ldots,\, 133$ 
& $E_7$ adjoint \\ 
$\alpha,\,\beta,\,\gamma$ & $1,\ldots,\, 56$
& $E_7$ fundamental \\
$s,\,t,\,u$ & $1,\,2,\,3$ & $SU(2)$ adjoint \\
$i,\,j,\,k$ & $1,\,2$   & $SU(2)$ fundamental \\ \hline
\end{tabular}
\end{center}
\caption{Group indices}
\end{table}
The $E_8$ commutation relations
have the form \cite{OW}:
$$
[Y_\Sigma,\,Y_\Lambda]\,=\,
f_{\Sigma\Lambda\Pi}Y_\Pi\;\qquad
[I_s,\,I_t]\,=\,\ve_{stu}I_u\;\qquad 
[Y_\Sigma,\,I_s]\,=\,0
$$
$$
[Y_\Sigma,\,Q_{i\alpha}]\,=\,(T_\Sigma)^\beta{}_\alpha
Q_{i\beta}\;\qquad
[I_s,\,Q_{i\alpha}]\,=\,-\,{i\over 2}\,(\sigma_s)^j{}_i
Q_{j\alpha}
$$
\be
\label{e8com}
[Q_{i\alpha},\,Q_{j\beta}]\,=\,-\,(CT_\Sigma)_{\alpha\beta}
\ve_{ij} Y_\Sigma\,+\,{i\over 2}\,C_{\alpha\beta}
(\ve\sigma_s)_{ij} I_s
\ee
Here $(T_\Sigma)^\alpha{}_\beta$ are 
antihermitean $56\times 56$ $E_7$
matrices,  normalized as 
$tr(T_\Sigma T_\Lambda)=-6\delta_{\Sigma\Lambda}$, 
$[T_\Sigma,T_\Lambda]=f_{\Sigma\Lambda\Pi}T_\Pi$, 
$f_{\Sigma\Lambda\Pi}$ are $E_7$ structure constants, 
$C_{\alpha\beta}$ is
antisymmetric $E_7$-invariant matrix,
$C^{\alpha\beta}$ denotes $C^+=C^{-1}$. 
$(\sigma_s)^i{}_j$ are
Pauli matrices, $\ve^{ij}=\ve_{ij}=i\sigma_2$.
The normalization of 
generators is chosen in such way, that $E_8$ 
Cartan-Killing metric is 
$-30(\delta_{\Sigma\Lambda},
\delta_{st}, C_{\alpha\beta}\ve_{ij})$.
Due to $E_7$ Fierz identity \cite{Cvi}
$$
(T_\Sigma)^\gamma{}_\alpha
(T_\Sigma)^\delta{}_\beta\,=\,
-\,{1\over 4}\left(\,\delta_\alpha^\delta
\delta_\beta^\gamma\,
+\,C^{\gamma\delta}C_{\alpha\beta}\,+
\,C^{\gamma\lambda}C^{\delta\sigma}
d_{\alpha\beta\lambda\sigma}\,\right)
$$
the relations (\ref{e8com}) satisfy
the Jacobi identities.
$d_{\alpha\beta\lambda\sigma}$ is totally symmetric
$E_7$-invariant tensor.

The scalars of the hypermultiplet $\Phi^{i\alpha}$
are the coordinates of the coset manifold. 
They satisfy the reality condition:
$$
\bar{\Phi}_{i\alpha}\,\equiv
(\Phi^{i\alpha})^*\,=\,\ve_{ij}C_{\alpha\beta}
\Phi^{j\beta}
$$
The vielbein $V^{i\alpha}$ and spin-connections
$\Omega^\Sigma,\,\Omega^s$ of the coset manifold can be
constructed by means of the Maurier-Cartan form
(for differential geometry of coset manifolds see,
for example, \cite{CAF}):
$$
L^{-1}\,{\partial \over \partial \Phi^{i\alpha}}\,L\,
=\,2i\,V_{\un{i\alpha}}{}^{j\beta}Q_{j\beta}\,+\,
\Omega_{\un{i\alpha}}{}^\Sigma Y_\Sigma\,+\,
\Omega_{\un{i\alpha}}{}^s I_s
$$
\be
\label{maca}
\mbox{where}\qquad 
L\,=\,\exp\left(2i\Phi^{i\alpha}Q_{i\alpha}\right)\qquad
\mbox{is the coset representative,}
\ee 
underlying index $\un{i\alpha}$ 
is curved one. The multiplier $2i$ is introduced for
conventional normalization of the field $\Phi$ in the
lagrangian. Evaluating equation (\ref{maca})
we obtain the following expressions for the vielbein
and spin-connections:
$$
V_{\un{i\alpha}}{}^{j\beta}\,=\,\left( \,
{{\rm sh}\sqrt{M} \over \sqrt{M}}\, \right)\!{}^{j\beta}
{}_{i\alpha}
$$
\be
\label{vsc}
\Omega_{\un{i\alpha}}{}^\Sigma\,=\,-\,4\left( \,\bar{\Phi}
T_\Sigma \,{{\rm ch}\sqrt{M}-1 \over M}\, \right)
\!_{i\alpha}\qquad
\Omega_{\un{i\alpha}}{}^s\,=\,-\,2i\left( \,\bar{\Phi}
\sigma_s \,{{\rm ch}\sqrt{M}-1 \over M}\, \right)
\!_{i\alpha}
\ee
where $M$ is hermitean $112\times 112$ matrix:
\be
M^{j\beta}{}_{i\alpha}\,=\,-\,4\,
(T_\Sigma\Phi)^{j\beta}(\bar{\Phi}T_\Sigma)_{i\alpha}
\,+\,(\sigma_s\Phi)^{j\beta}(\bar{\Phi}\sigma_s)_{i\alpha}
\ee
In particular $M^{i\alpha}{}_{j\beta}\Phi^{j\beta}=0$.
Consequently the matrix $M$ is degenerate and doesn't have 
inverse one, so the equations (\ref{vsc}) should be
considered as formal expansions in powers of $M$, which
are positive. Due to the properties
$$
\partial_{\un{k\gamma}}f(M)^{j\beta}{}_{i\alpha}
\,(T_\Sigma\Phi)^{k\gamma}\,=\,
[T_\Sigma,f(M)]^{j\beta}{}_{i\alpha}
$$
\be
\label{pr1}
\partial_{\un{k\gamma}}f(M)^{j\beta}{}_{i\alpha}
\,(\sigma_s\Phi)^{k\gamma}\,=\,
[\sigma_s,f(M)]^{j\beta}{}_{i\alpha}
\ee
any function $f(M)$ transforms uniformly 
under $E_7\times SU(2)$ transformations:
$$
\delta\Phi\,=\,U^\Sigma T_\Sigma\Phi\,
\,-\,{i\over 2}U^s\sigma_s\Phi
$$
Furthermore, the calculation of 
$\partial_{[i\alpha}(L^{-1}\partial_{j\beta]}L)$
gives the following equations on the derivatives
of the vielbein and spin-connections:
$$
\partial_{[\un{i\alpha}}V_{\un{j\beta}]}{}^{k\gamma}\,
+\,(T_\Sigma)^\gamma{}_\delta
\Omega_{[\un{i\alpha}}{}^\Sigma
V_{\un{j\beta}]}{}^{k\delta}\,-\,
{i \over 2}\,(\sigma_s)^k{}_l\Omega_{[\un{i\alpha}}{}^s
V_{\un{j\beta}]}{}^{l\gamma}\,=\,0
$$
$$
\partial_{[\un{i\alpha}}\Omega_{\un{j\beta}]}{}^\Sigma\,+\,
{1 \over 2}\,f_{\Sigma\Lambda\Pi}
\Omega_{\un{i\alpha}}{}^\Lambda
\Omega_{\un{j\beta}}{}^\Pi\,+\,2\,
(CT_\Sigma)_{\gamma\delta}\ve_{kl}
V_{\un{i\alpha}}{}^{k\gamma}V_{\un{j\beta}}{}^{l\delta}
\,=\,0
$$
\be
\label{pr2}
\partial_{[\un{i\alpha}}\Omega_{\un{j\beta}]}{}^s\,+\,
{1 \over 2}\,\ve_{stu}\Omega_{\un{i\alpha}}{}^t
\Omega_{\un{j\beta}}{}^u\,-\,i\,
C_{\gamma\delta}(\ve\sigma_s)_{kl}
V_{\un{i\alpha}}{}^{k\gamma}V_{\un{j\beta}}{}^{l\delta}
\,=\,0
\ee
We will use the properties (\ref{pr1}), (\ref{pr2})
evaluating the supersymmetry algebra.

\section*{Coupling to supergravity}

We construct the lagrangian following Noether
procedure. One starts with the lagrangian of pure
N=2, D=6 supergravity and adds terms, necessary to 
cancel its variation with respect to the
supersymmetry transformations, modified due to the
presence of the matter fields. The algebra
is very similar to the $Sp(n)$ case \cite{NS},
so we omit details of the calculation. We consider
dual version of the N=2, D=6 supergravity, so
that the field-strength tensor 
$B_{mnp}=3\partial_{[m}B_{np]}$ does not contain
Chern-Simons term. The lagrangian of the dual
supergravity coupled to $Sp(n)$-matter can be found
in \cite{NG}. For simplicity we omit fourth-order
fermionic terms in the lagrangian and third-order
fermionic terms in the supersymmetry transformations of
the fermions.

Letters $a,b,c$ are used as flat space-time indices 
and $m,n,p$ as curved ones. 
The metric and antisymmetric tensors are 
$\eta_{ab}=(+,-,\ldots,-)$, $\ve^{01\ldots 5}=1$. 
We use 8-component spinors and $8\times 8$ Dirac matrices
$\gamma^a$. Spinorial indices are not written explicitly 
in the text.
As usual, $\gamma^{a_1\ldots a_n}$ denotes
antisymmetrized product of $n$ $\gamma$-matrices. Notations
of all fields and their conjugation rules are given in
Table 2.
\begin{table}[bt]
\begin{center}
\begin{tabular}{llccl}
\hline
multiplet: & fields: & notations: & 
{\small $E_7\times SU(2)$} rep.:  
& reality: \\ \hline\hline
gravitational + & graviton & $e_m{}^a$ & (1,1) & real \\
antisymmetric: & antisym. tensor &$B_{mn}$& (1,1) & real \\
 & dilaton & $\varphi$ & (1,1) & real \\
 & gravitino &$\psi_m^i$& (1,2) & 
$\tilde{\psi}_{mi}=i\ve_{ij}\psi^j_m$   \\
 & dilatino &$\cchi^i$& (1,2) & 
$\tilde{\cchi}_i=-i\ve_{ij}\cchi^j$ \\
\hline
hypermultiplet: & scalar & $\Phi^{i\alpha}$ & (56,2) &
$\bar{\Phi}_{i\alpha}=
\ve_{ij}C_{\alpha\beta}\Phi^{j\beta}$\\
 & fermion & $\Psi^\alpha$ & (56,1) &
$\tilde{\Psi}_\alpha=-iC_{\alpha\beta}\Psi^\beta$ \\
\hline
$E_7$ vector: & vector & $A_m^\Sigma$ & (133,1) & real \\
 & fermion & $\lambda^{\Sigma i}$ & (133,2) &
$\tilde{\lambda}{}^\Sigma_i=i\ve_{ij}\lambda^{\Sigma j}$ \\
\hline
$SU(2)$ vector: & vector & ${\cal A}_m^s$ & (1,3) 
& real \\
 & fermion & $\rho^{si}$ & (1,3$\times$2) &
$\tilde{\rho}{}^s_i=i\ve_{ij}\rho^{sj}$ \\
\hline 
\end{tabular}
\end{center}
\caption{N=2, D=6 supersymmetric multiplets. 
For all spinors $\tilde{\psi}=\bar{\psi}{\cal C}^+$,
$\bar{\psi}=\psi^*\gamma^0$, ${\cal C}$ is unitary
symmetric charge
conjugation matrix. Dilatino and $\Psi$ are right-handed
spinors $\gamma^7\chi =-\chi$, all other fermions are
left-handed.}
\end{table}
The vielbein
$V_{i\alpha}{}^{\un{j\beta}}$ is inverse to
$V_{\un{i\alpha}}{}^{j\beta}$ one and 
$g_{i\alpha,j\beta}=\ve_{kl}C_{\gamma\delta}
V_{\un{i\alpha}}{}^{k\gamma}V_{\un{j\beta}}{}^{l\delta}$
is the metric of the coset manifold.

At first let us write down the definitions of 
the covariant derivatives for all fields:
\bea
\label{der}
D_m\Phi & = & \partial_m\Phi\,-\,A_m^\Sigma T_\Sigma\Phi\,
+\,{i\over 2}\,{\cal A}_m^s\sigma_s\Phi \nonumber \\
D_m\Psi & = &\Psi_{;\,m}\,-\,A_m^\Sigma T_\Sigma\Psi\,
+\,D_m\Phi^{i\alpha}\Omega_{\un{i\alpha}}{}^\Sigma
T_\Sigma\Psi \nonumber \\
D_m\epsilon & = & \epsilon_{;\,m}\,
+\,{i \over 2}\,{\cal A}_m^s\sigma_s\epsilon 
\,-\,{i \over 2}\,D_m\Phi^{i\alpha}
\Omega_{\un{i\alpha}}{}^s\sigma_s\epsilon\qquad
\mbox{the same for}\quad \psi_m,\,\cchi \nonumber \\
D_m\lambda^\Sigma & = &\lambda_{;\,m}^\Sigma
\,-\,f_{\Sigma\Lambda\Pi} 
A_m^\Lambda \lambda^\Pi\,
+\,{i\over 2}\,{\cal A}_m^s\sigma_s\lambda^\Sigma\,-\,
{i \over 2}\,D_m\Phi^{i\alpha}
\Omega_{\un{i\alpha}}{}^s\sigma_s\lambda^\Sigma
\nonumber \\
D_m\rho^s & = &\rho_{;\,m}^s\,-\,
\ve_{stu}{\cal A}_m^t\rho^u\,
+\,{i\over 2}\,{\cal A}_m^t\sigma_t\rho^s\,-\,
{i \over 2}\,D_m\Phi^{i\alpha}
\Omega_{\un{i\alpha}}{}^t\sigma_t\rho^s
\eea
The semicolon denotes usual space-time covariant 
derivative; $\epsilon$ is the parameter of the 
supersymmetry transformations. The field-strength
tensors are:
\bea
F^\Sigma_{mn}&=&2\,\partial^{}_{[m}A^\Sigma_{n]}\,-\,
f_{\Sigma\Lambda\Pi}A_m^\Lambda A_n^\Pi
\nonumber \\
{\cal F}^s_{mn}&=&2\,
\partial^{}_{[m}{\cal A}^s_{n]}\,-\,
\ve_{stu}{\cal A}_m^t{\cal A}_n^u
\eea
According to \cite{NS}, the derivatives of the
fields, belonging to fundamental representations
of $E_7\times SU(2)$ get modified by adding the
terms with the spin-connections $\Omega$.
So the commutator becomes:
$$
D_{[m}D_{n]}\epsilon\,=\,-\,{1 \over 8}\,R_{mncd}
\gamma^{cd}\epsilon\,+\,{i \over 4}\,{\cal F}_{mn}{}^s
\sigma_s\epsilon\,+
$$
$$
+\,{i \over 4}\,\sigma_s\epsilon\,
\Omega_{\un{i\alpha}}{}^s\left(\,F_{mn}^\Sigma T_\Sigma\Phi
\,-\,{i \over 2}\,{\cal F}_{mn}^s\sigma_s\Phi\,
\right)\!{}^{i\alpha} +\,{1 \over 2}\,\sigma_s\epsilon\,
D_m\Phi^{i\alpha}D_n\Phi^{j\beta}
V_{\un{i\alpha}}{}^{k\gamma}V_{\un{j\beta}}{}^{l\delta}
C_{\gamma\delta}\,(\ve\sigma_s)_{kl}
$$
We used properties (\ref{pr1}), (\ref{pr2})
proving this.

The lagrangian of the N=2, D=6 dual supergravity
coupled with $E_7\times SU(2)$ vector multiplets and
hypermultiplet has the following form:
\bea
\label{lagr}
e^{-1}L & = & {1\over 4}\,R\,+\,
\varphi_{;\,a}\varphi^{;\,a}\,
+\,{1\over 12}\,e^{4\varphi} B_{abc}B^{abc}\,
-\,{i\over 2}\,\bar{\psi}_a\gamma^{abc}D_b\psi_c\,
+\,{i\over 2}\,\bar{\cchi}\hat{D}\cchi\,
\nonumber \\ 
& &
-\,i\,\varphi_{;\,a}\bar{\psi}_b\gamma^a\gamma^b\cchi\,
+\,{i \over 24}\,e^{2\varphi}\,\left(\,
\bar{\psi}_a\gamma^{[a}{\hat B}\gamma^{b]}\psi_b\,
+\,2\,\bar{\psi}_a\hat{B}\gamma^a\cchi\,
+\,\,\bar{\cchi}{\hat B}\cchi\,
-\,\bar{\Psi}\hat{B}\Psi\,\right)
\nonumber \\
& &
+\,{1 \over 2}\,g_{i\alpha,j\beta}D_a\Phi^{i\alpha}
D^a\Phi^{j\beta}
\,+\,{i \over 2}\,\bar{\Psi}\hat{D}\Psi\,
-\,i\,D_a\Phi^{i\alpha}V_{\un{i\alpha}}{}^{j\beta}
\ve_{jk}(\bar{\Psi}_\beta\gamma^b\gamma^a\psi^k_b)
\nonumber \\
& &
+\,{1 \over g^2}\left[
\,-\,{1\over 4}\,e^{-2\varphi}
F_{ab}^\Sigma F^{\Sigma\,ab}\,
+\,{1\over 8}\,\ve^{a_1 \ldots a_6}
F_{a_1 a_2}^\Sigma F_{a_3 a_4}^\Sigma B^{}_{a_5 a_6}\,
+\,i\,e^{-2\varphi}\bar{\lambda}{}^\Sigma
\hat{D}\lambda^\Sigma
\right.
\nonumber \\
& &
\left.
+\,{i\over 2}\,e^{-2\varphi}\bar{\lambda}{}^\Sigma
\left(\gamma^a\hat{F}\psi_a\,
+\,\hat{F}\cchi\right)\!{}^\Sigma\,+
\,{i\over 12}\,\bar{\lambda}{}^\Sigma\hat{B}\lambda^\Sigma
\,\right]
-\,2i\,(T_\Sigma\Phi)^{i\alpha}
V_{\un{i\alpha}}{}^{j\beta}
\ve_{jk}(\bar{\Psi}_\beta\lambda^{\Sigma i})
\nonumber \\
& &
+\,{1 \over {g'}^2}\left[
\,-\,{1\over 4}\,e^{-2\varphi}
{\cal F}_{ab}^s {\cal F}^{s\,ab}\,
+\,{1\over 8}\,\ve^{a_1 \ldots a_6}
{\cal F}_{a_1 a_2}^s {\cal F}_{a_3 a_4}^s B^{}_{a_5 a_6}\,
+\,i\,e^{-2\varphi}\bar{\rho}{}^s\hat{D}\rho^s
\right.
\nonumber \\
& &
\left.
+\,{i\over 2}\,e^{-2\varphi}\bar{\rho}{}^s
\left(\gamma^a\hat{\cal F}\psi_a\,
+\,\hat{\cal F}\cchi\right)\!{}^s\,+
\,{i\over 12}\,\bar{\rho}{}^s\hat{B}\rho^s
\,\right]
-\,(\sigma_s\Phi)^{i\alpha}
V_{\un{i\alpha}}{}^{j\beta}\ve_{jk}(\bar{\Psi}_\beta
\rho^{si})
\nonumber \\
& &
+\,{1\over 2}\,(\,\bar{\psi}_a\gamma^a\,+\,\bar{\cchi}\,)\,
\sigma_s\,(\,\lambda^\Sigma C^{\Sigma s}\,+\,
\rho^t C^{ts}\,)
\,-\,{1\over 8}\,e^{2\varphi}(g^2 C^{\Sigma s}
C^{\Sigma s}+{g'}^2 C^{st}C^{st})
\eea
where $\hat{B}=B^{abc}\gamma_{abc}$ and so on;
$g,\,g'$ are $E_7$ and $SU(2)$ coupling constants
respectively. Following \cite{NS}, we introduced
real functions
$$
C^{\Sigma s}\,=\,-\,2i\,\bar{\Phi}\sigma_s\,
{{\rm ch}\sqrt{M}-1 \over M}\,T_\Sigma\Phi\qquad
C^{st}\,=\,\delta^{st}\,-\,
\bar{\Phi}\sigma_t\,{{\rm ch}\sqrt{M}-1 \over M}
\,\sigma_s\Phi
$$
for notational convenience. For the same reason
indices, labeling representations 2 and 56 are
suppressed in those places, where their position
can be restored unambiguously.

The lagrangian (\ref{lagr}) is invariant with respect to
the supersymmetry transformations, written below:
\bea
\label{susytr}
\delta e_m{}^a & = & i\,\bar{\psi}_m\gamma^a\epsilon
\nonumber \\
\delta\varphi & = & -\,{i\over 2}\,\bar{\cchi}\epsilon
\nonumber \\
\delta B_{mn} & = & i\,e^{-2\varphi}\,\left(\,
-\,\bar{\psi}_{[m}\gamma_{n]}\epsilon\,
-\,{1\over 2}\,\bar{\cchi}\gamma_{mn}\epsilon\,\right)
\nonumber \\
\delta\psi_m & = & D_m\epsilon\,
-\,{1\over 24}\,e^{2\varphi}\hat{B}\gamma_m\epsilon\,
\nonumber \\
\delta\cchi & = & -\,\varphi_{;\,a}\gamma^a\epsilon
-\,{1\over 12}\,e^{2\varphi}\hat{B}\epsilon
\nonumber \\
\delta\Phi^{i\alpha} & = & i\,
V_{j\beta}{}^{\un{i\alpha}}\,
C^{\beta\gamma}\,(\bar{\Psi}_\gamma\epsilon^j)
\nonumber \\
\delta\Psi^\alpha & = & D_a\Phi^{i\beta}
V_{\un{i\beta}}{}^{j\alpha}\ve_{jk}\gamma^a\epsilon^k
\nonumber \\
\delta A_m^\Sigma & = & i\,\bar{\lambda}{}^\Sigma
\gamma_m\epsilon
\nonumber \\
\delta \lambda^\Sigma & = & -\,{1 \over 4}\,
\hat{F}{}^\Sigma\epsilon\,
+\,{i\over 4}\,g^2e^{2\varphi}C^{\Sigma s}\sigma_s\epsilon
\nonumber \\
\delta {\cal A}_m^s & = & i\,\bar{\rho}{}^s\gamma_m\epsilon
\nonumber \\
\delta \rho^s &=& -\,{1 \over 4}\,\hat{\cal F}{}^s
\epsilon\,+\,{i\over 4}
\,{g'}^2e^{2\varphi}C^{st}\sigma_t\epsilon
\eea

The lagrangian (\ref{lagr}) has the same form as
in the case of $Sp(n)$-matter \cite{NG}.
Nevertheless there is one essential difference.
In our case there is no need to impose the
constraint 
$\ve_{kl}V_{(\un{i\alpha}}{}^{k\gamma}
V_{\un{j\beta})}{}^{l\delta}=-n^{-1}
g_{i\alpha,j\beta}C^{\gamma\delta}$ on the vielbein.
This constraint has been introduced in \cite{BW}
(eq.(2)) and used in \cite{NS,NG}. We never used
this constraint checking the invariance of the lagrangian
(\ref{lagr}) with respect to the transformations 
(\ref{susytr}). Moreover, the vielbein (\ref{vsc})
does not satisfy this constraint. It can be easily
seen in the point $\Phi=0$ where
$V_{\un{i\alpha}}{}^{j\beta}
=\delta_i^j\delta_\alpha^\beta$.

\section*{Acknowledgements}

Author would like to express his debt to M.V.Terentiev
for many guiding ideas on the early stage of this work.
Author thanks faculty of the University of Michigan
for kind hospitality.


\begin{thebibliography}{199}
\bibitem{NS}
H.Nishino, E.Sezgin, Phys.Lett. B144 (1984) 187;
Nucl.Phys. B278 (1986) 353; hep-th/9703075
\bibitem{BW}
J.Bagger, E.Witten, Nucl.Phys. B222 (1983) 1
\bibitem{CHSW}
P.Candelas, G.Horowitz, A.Strominger, E.Witten,
Nucl.Phys. B258 (1985) 46
\bibitem{GSW}
M.B.Green, J.H.Schwarz, P.C.West, 
Nucl.Phys. B254 (1985) 327 
\bibitem{TZ}
M.Terentiev, K.Zyablyuk, hep-th/9608119
\bibitem{FKP}
S.Ferrara, C.Kounnas, M.Porrati, Phys.Lett. B181 (1986) 263
\bibitem{SS}
A.Salam, E.Sezgin, Phys.Lett. B147 (1984) 47
\bibitem{OW}
D.Olive, P.West, Nucl.Phys. B217 (1983) 248
\bibitem{Cvi}
P.Cvitanovi\'c, Phys.Rev. D14 (1976) 1536
\bibitem{CAF}
L.Castellani, R.D'Auria, P.Fr\'e, 
{\it Supergravity and Superstrings -- A Geometric
Perspective}, Vol.1; World Scientific, 1991
\bibitem{NG}
H.Nishino, S.J.Gates Jr., Nucl.Phys. B268 (1986) 532
\end{thebibliography}
\end{document}